\documentclass[aps,prd,twocolumn,groupedaddress,showpacs,amsmath,amssymb]{revtex4-1}

\usepackage{graphicx}
\usepackage{bm,color}

\newcommand{\be}{\begin{eqnarray}}
\newcommand{\ee}{\end{eqnarray}}
\newcommand{\nn}{\nonumber\\}

\def\tQ{\tau_{\rm Q}}
\def\teq{t_{\rm eq}}
\def\tex{t_{\rm ex}}
\def\hx{\hat{x}}
\def\hy{\hat{y}}

\begin{document}

\title{Dynamics of a quantum phase transition in the
Bose-Hubbard model: \\
Kibble-Zurek mechanism and beyond
}
\author{Keita Shimizu$^1$}
\author{Yoshihito Kuno$^2$}
\author{Takahiro Hirano$^1$}
\author{Ikuo Ichinose$^1$}
\affiliation{$^1$Department of Applied Physics, Nagoya Institute of Technology, Nagoya, 466-8555, Japan}
\affiliation{$^2$Department of Physics, Graduate School of Science, Kyoto University, 
Kyoto, 606-8502, Japan}

\date{\today}

\begin{abstract}
In this paper, we study the dynamics of the Bose-Hubbard model by using
time-dependent Gutzwiller methods.
In particular, we vary the parameters in the Hamiltonian
as a function of time, 
and investigate the temporal behavior of the system from
the Mott insulator to the superfluid (SF) crossing a second-order phase transition.
We first solve a time-dependent Schr\"{o}dinger equation for the experimental
setup recently done by Braun {\it et.al.} 
[Proc. Nat. Acad. Sci. {\bf 112}, 3641 (2015)] and 
show that the numerical and experimental results are in fairly good agreement.
However, these results disagree with the Kibble-Zurek scaling.
From our numerical study, we reveal a possible source of the discrepancy.
Next, we calculate the critical exponents of the correlation length and vortex density
in addition to the SF order parameter for a Kibble-Zurek protocol.
We show that beside the ``freeze'' time $\hat{t}$, there exists another
important time, $t_{\rm eq}$, at which an oscillating behavior of the SF
amplitude starts. 
From calculations of the exponents of the correlation length and vortex density
with respect to a quench time $\tQ$, we obtain a physical picture of 
a coarsening process.
Finally, we study how the system evolves after the quench.
We give a global picture of dynamics of the Bose-Hubbard model.
\end{abstract}

\pacs{
67.85.Hj,	
03.75.Kk,	
05.30.Rt	
}
\maketitle


\section{Introduction}{\label{intro}}

In recent years, ultra-cold atomic gas systems play an very important role
for the study on the quantum many systems because of
their versatility and controllability.
Sometimes they are called quantum simulators \cite{Nori,Cirac,coldatom1,coldatom2}.
In this paper, we focus on the dynamical behavior of many-body systems 
for which ultra-cold Bose gas systems are feasible as a quantum 
simulator.

The problem how a system evolves under a change in temperature
crossing a second (continuous) thermal phase transition has been studied
extensively.
For this problem, Kibble \cite{kibble1,kibble2} pointed out from the view point of 
the cosmology that the system exhibits non-equilibrium behavior
and the phase transitions lead to disparate local choices of the broken
symmetry vacuum and as a result, topological defects are generated.
Later, Zurek \cite{zurek1,zurek2,zurek3} suggested that a similar phenomenon 
can be realized in experiments on
the condensed matter systems like the superfluid (SF) of $^4$He.
After the above seminal works by Kibble and Zurek, there appeared many
theoretical and experimental studies on the Kibble-Zurek mechanism (KZM)~\cite{IJMPA}.
Experiments with ultra-cold atomic gases were done to verify the KZ scaling law
for exponents of the correlation length and the rate of topological defect formation
with respect to the quench time \cite{navon}.

In this paper,  we shall study how low-energy states evolve under a ramp change of
the parameters in the Hamiltonian crossing a quantum phase transition (QPT),
i.e., the quantum quench
\cite{Chomaz,dziarmaga,pol,Zoller,sondhi,francuz,Be1,Be2,Zu1,Zu2,KZMII}.
This problem is also attracted great interest in recent years, and experiments 
investigating dynamics of many-body quantum systems through QPTs were already done 
using the ultra-cold atomic gases as a quantum 
simulator~\cite{Chen,Braun,Anquez,clark}.
Works in Refs.~\cite{Chen,Braun} questioned the applicability of the KZ scaling
theory to the QPT, whereas Ref.~\cite{Anquez} used a spin-1 Bose-Einstein
condensation (BEC) with a small size in order to avoid domain formation and concluded
that the satisfactory agreement of the measured scaling exponent with the KZ theory
was obtained.
In Ref.~\cite{clark}, BECs in a shaken optical lattice were studied.
There, BECs exhibit a phase transition from a ferromagnetic to anti-ferromagnetic
spatial correlations, and the observed results were in good agreement with 
the KZ scaling law.
The above facts indicate the necessity of the further studies on the dynamics
of quantum many-body  systems.
In this paper, we focus on the two-dimensional (2D) Bose-Hubbard model 
(BHM)~\cite{BHM1,BHM2}, 
which is a canonical model of the bosonic ultra-cold atomic gas systems.
In particular, we investigate the out-of-equilibrium evolution of the system
under the linear ramp.
In our study, the system starts from equilibrated Mott insulator and passes 
through the equilibrium phase transition point to the SF.
From the experimental point of view, this is an ideal setup to observe dynamics of 
a quantum phase transition and to test Kibble-Zurek conjecture
in ultra-cold atomic gases.

This paper is organized as follows.
In Sec.~\ref{BHM}, we introduce the BHM and explain the protocol of the quench.
The time-dependent Gutzwiller (tGW) methods 
\cite{tGW1,tGW2,tGW3,tGW4,tGW5,tGW6,aoki} are briefly explained.

In Sec.~\ref{exper}, we show the numerical results obtained for the quench protocol
employed in the experiment in Ref.~\cite{Braun}.
Concerning to the exponent of the correlation length, the experiments and
our calculations are in fairly good agreement, whereas the KZ scaling theory using
the 3D XY model exponents disagrees with these measurements.
We reveal a possible source of this discrepancy by examining the behavior of the BEC 
as a function of time.

Section~\ref{KZMP} gives detailed study on the BHM for the KZ original 
protocol \cite{IJMPA}.
We first exhibit the temporal behavior of the BEC amplitude, and show the snapshots of
the local density and phase of BEC and also the vortex configuration.
From these observations, we point out the existence of another important time,
$\teq$, besides the ``freeze" time $\hat{t}$.
Calculations of the exponents of the correlation length, $\xi$, and vortex density, 
$N_{\rm v}$, at $t=\hat{t}$ and $t=\teq$ indicate that rather smooth evolution of the
system takes place from $\hat{t}$ to $\teq$.
In this regime, typical time scale and typical length scale are given by
$\hat{t}$ and the correlation length $\xi(\hat{t})$, respectively.
After $\teq$, a substantial coarsening process takes place that accompanies
oscillation of the density of the BEC.
Physical picture of the evolution of the system in the quench is given.

We introduce the time $\tex$ at which the above mentioned density oscillation of 
the BEC terminates.
In Sec.~\ref{after}, we investigate how the system evolves after $\tex$, 
in particular, if the system settle down to an equilibrium state.

Section \ref{conclusion} is devoted to conclusion and discussion.
In particular, we discuss the experimental setup to observe phenomena
studied in the work.

\section{Bose-Hubbard model and slow quench}{\label{BHM}}

We shall consider the Bose-Hubbard model Hamiltonian of which is given 
as follows,
\be
H_{\rm BH}&=&-J\sum_{\langle i,j \rangle}(a^\dagger_i a_j+\mbox{H.c.})
+{U \over 2}\sum_in_i(n_i-1)  \nn
&&-\mu\sum_in_i,
\label{HBH}
\ee
where $\langle i, j\rangle$ denotes nearest-neighbor (NN) sites,
$a_i^\dagger \ (a_i)$ is the creation (annihilation) operator of boson at site $i$,
$n_i=a^\dagger_i a_i$, and $\mu$ is the chemical potential.
$J(>0)$ and $U(>0)$ are the hopping amplitude and the on-site repulsion, respectively.
For $J\ll U$, the system is in a Mott insulator, whereas for $J \gg U$, the SF forms.
For the 2D system at the unit filling, we estimated the critical value 
as $(J/U)_{\rm c}=0.043$ by using the static Gutzwiller methods.
In the present paper, we consider these parameters are time-dependent as they are to
be controlled in the experiments, i.e., $J=J(t)$ and $U=U(t)$.

Detailed protocols will be explained in the subsequent section in which
calculations are given.
However, a typical protocol of the KZM employed in this work is the following;
\begin{enumerate}
\item In the practical calculation, we fix the value of $U$ as the unit of energy.
[$U=1$.]
\item Time $t$ is measured in the unit $\hbar/U$.
\item The hopping amplitude is varied as 
\be
{J(t)-J_c \over J_c}\equiv \epsilon(t)={t \over \tau_{\rm Q}},
\label{protocol}
\ee
where $J_c$ is the critical hopping for the fixed $U$ and $\tau_{\rm Q}$
is the quench time, which is a controllable parameter in experiments.
\end{enumerate}
We solve a time-dependent Schr\"{o}dinger equation for the system
$H_{\rm BH}$ with $J(t)$ from $t=t_{\rm i}(<0)$ to $t=t_{\rm f}(>0)$.
At $t=t_{\rm i}$, the system is in the Mott insulator phase. 
On the other hand, at $t=t_{\rm f}$, the resulting system is in the SF parameter region.
The hopping $J(t)$ is linearly increased with time and approaches to the critical 
value $J_c$ at $t=0$.
However, because of the rapid increase of the relaxation time, $t_{\rm r}$, 
the system cannot follow the change in $J(t)$, and the adiabatic process is terminated 
at $t=-\hat{t}$. 
As the relaxation time is given as 
$t_{\rm r}=\epsilon(t)^{-z\nu} \equiv |{J(t)-J_c \over J_c}|^{-z\nu}
=|{t \over \tau_{\rm Q}}|^{-z\nu}$, the condition $t_{\rm r}=\hat{t}$ gives
$\hat{t}=(\tau_{\rm Q})^{z\nu/(1+z\nu)}$, where $\nu$ is the critical exponent
of the correlation length and $z$ is the dynamical critical exponent.
The relaxation time then decreases after passing through the phase transition 
point and then the system reenters the ``adiabatic regime".
The regime between $(-\hat{t}, \hat{t})$ is sometimes called a ``frozen" region,
but the system evolves slowly compared with the relaxation time even in the 
frozen regime.

The KZM predicts the scaling law of the correlation length $\xi(t)$ and 
the vortex density $N_{\rm v}(t)$ at $t=\hat{t}$ as
$\xi(\hat{t})\propto (\tau_{\rm Q})^{\nu/(1+z\nu)}$ and
$N_{\rm v}(\hat{t})\propto (\tau_{\rm Q})^{-2\nu/(1+z\nu)}$.
We shall investigate the above KZ scaling law in the present work by using
the tGW methods.
We mostly focus on the unit filling case although to study other fillings
is straightforward.

We employ the tGW methods for the calculation 
\cite{tGW1,tGW2,tGW3,tGW4,tGW5,tGW6,aoki}, which 
are efficient for numerical simulations of the real-time dynamics 
of 2D or higher dimensional systems. 
During time evolution, the particle-number conservation is satisfied by using small enough
(dimensionless) time step $dt$ ($\mathcal{O}(dt)\lesssim 10^{-4}$). 
Similarly for isolated systems, 
the tGW methods satisfy the total energy conversation during time evolutions.

In tGW approximation, the Hamiltonian of the BH model [Eq.(\ref{HBH})]
is devised into a single-site Hamiltonian $H_i$ by introducing the expectation 
value $\Psi_i=\langle a_i \rangle$,
\be 
&&H_{\rm GW}=\sum_i H_i,   \nn
&&H_i=-J\sum_{j\in i{\rm NN}}(a^\dagger_i\Psi_j+\mbox{H.c.})
+{U \over 2}\sum_in_i(n_i-1)  \nn
&&\hspace{1cm}-\mu\sum_in_i,
\label{HGW}
\ee
where $i{\rm NN}$ denotes the NN sites of site $i$.
Then, the GW wave function is given as,
\be 
|\Phi_{\rm GW}\rangle=\prod^{N_s}_i\Big(\sum^{n_c}_{n=0}
f^i_n(t)|n\rangle_i\Big),
\;\; \hat{n}_i|n\rangle_i=n|n\rangle_i,
\ee 
where $N_s$ is the number of the lattice sites and $n_c$ is the maximum 
number of particle at each site.
For the study on the unit filling case, $\langle n_i \rangle=1$,
we verified that $n_c=6$ is large enough.
In terms of $\{f^i_n(t)\}$, the order parameter of the BEC is given as,
\be
\Psi_i=\langle a_i \rangle=\sum^{n_c}_{n=0}\sqrt{n}f^{i\ast}_{n-1}f^i_n,
\label{BEC}
\ee
and $\{f^i_n(t)\}$ are determined by solving the following Schr\"{o}dinger equation
for various initial states corresponding to the Mott insulator,
\be i\hbar \partial_t |\Phi_{\rm GW}\rangle
=H_{\rm GW}(t)|\Phi_{\rm GW}\rangle.
\label{SEq}
\ee
The time dependence of $H_{\rm GW}(t)$ in Eq.(\ref{SEq}) stems from the quench
$J\to J(t)$.
For solving the Schr\"{o}dinger equation (\ref{SEq}), we use the fourth-order 
Runge-Kutta methods \cite{Fortran}.
We have verified that the total energy of the system is conserved for the case of the
constant values of $J$ and $U$.
In the calculations in the following sections, observables are averaged at fixed $t$ over 10 initial configurations \cite{initial}.


\section{Numerical study for experiment protocol}{\label{exper}}

\begin{figure}[t]
\centering
\begin{center}
\includegraphics[width=7cm]{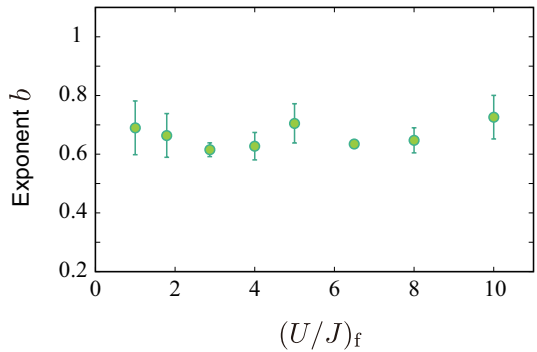}
\end{center}
\caption{Exponents of the correlation length $b$ for the protocol
in Ref.~\cite{Braun}.
The results are in fairly good agreement with the measurements reported in 
Ref.~\cite{Braun}.
However, they do not agrees with the value predicted by the KZ scaling hypothesis.
}
\label{PNAS}
\end{figure}
\begin{figure}[t]
\centering
\begin{center}
\includegraphics[width=6.5cm]{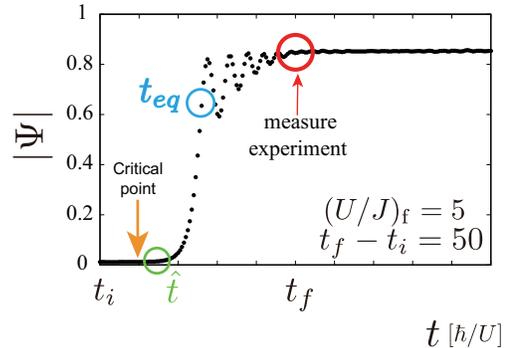}
\end{center}
\caption{Temporal evolution of the amplitude of the order parameter
$|\Psi|$ for the protocol in Ref.~\cite{Braun}.
Critical point, $\hat{t}$ and time of measurement in Ref.~\cite{Braun} are
indicated.
}
\label{OPPNAS}
\end{figure}

Recently, experiments on the {\em quantum} dynamical behavior of ultra-cold atomic
systems were performed.
Among them, the experiment by Braun et al. measured the exponent of the correlation
length $b$, $\xi \propto (\tau_{\rm Q})^b$, in one, two and three-dimensional systems
\cite{Braun}.
Their protocol is the following;
the system at first locates in the Mott insulator region with certain initial value
$(U/J)_{\rm i}$.
Then $(U/J)$ is decreased passing through the critical value $(U/J)_c$
and the quench is terminated in the SF region with $(U/J)_{\rm f}$.
They call $\tau_{\rm Q}=t_{\rm f}-t_{\rm i}$, where $t_{\rm i} \ (t_{\rm f})$
is the starting (ending) time of the quench.
In the experiment, $(U/J)_{\rm i}/(U/J)_{\rm f}\approx 35$.

We calculate the physical quantities such as the order parameter of the BEC,
correlation length $\xi$, vortex number $N_{\rm v}$, etc.
The correlation length $\xi$ and vortex number $N_{\rm v}$ are defined as follows,
\begin{eqnarray}
&&\langle \Psi_i^\ast \Psi_j\rangle \propto \exp (-|i-j|/\xi), \nonumber \\
&&N_{\rm v}=\sum_i|\Omega_i|, \nonumber \\
&&\Omega_i={1 \over 4}\Big[\sin (\theta_{i+\hx}-\theta_i)
+\sin (\theta_{i+\hx+\hy}-\theta_{i+\hx})
 \nonumber \\
&&\hspace{1cm} -\sin (\theta_{i+\hx+\hy}-\theta_{i+\hy})
-\sin (\theta_{i+\hy}-\theta_{i})\Big],
\end{eqnarray}
where $\theta_i$ is the phase of $\Psi_i$
and $\hx \ (\hy)$ is the unit vector in the $x \ (y)$ direction.
Adopting the experimental setup in Ref.~\cite{Braun} for the protocol,
we calculated the correlation length for various values of $(U/J)_{\rm f}$, and 
the obtained results for the correlation-length exponent $b$ are shown in Fig.~\ref{PNAS}.
We used 10 samples as the initial state of the Mott insulator, in which
particle numbers at sites have small fluctuations whereas relative phases
of $\{f^i_n(t_{\rm i})\}$ fluctuate strongly.
Fig.~\ref{PNAS} shows that the exponent $b$ is almost constant for various values of
$(U/J)_{\rm f}$, i.e., $b\simeq 0.6-0.7$.
This result is in fairly good agreement with the measurements of the experiment
for the 2D system \cite{Braun}
although there is a small but finite increase of $b$ for $(U/J)_{\rm f}=2.0-3.0$ 
in the experimental measurements. 
In Ref.~\cite{Braun}, the measured result was compared with the value of the KZ
scaling, i.e., $b=\nu/(1+z\nu)$.
The 3D XY model has the exponents $\nu=0.672$ and $z=1$ \cite{XY}, and therefore
KZ scaling predicts $b=0.402$.
It is obvious that the above value disagrees with the measurements.
From this fact, it was concluded that the KZ scaling and KZM could not
be applicable to the 2D quantum phase transition in Ref.~\cite{Braun}.  

The above results should be carefully examined to find a possible source of
the discrepancy.
To this end, we calculated the average amplitude of the BEC,
$|\Psi|={1 \over N_s}\sum_i|\Psi_i|$ as a function of time.
The result is shown in Fig.~\ref{OPPNAS}.
$|\Psi|$ exhibits an interesting behavior as first observed in Ref.~\cite{aoki}
for the {\em sudden quench}.
After crossing the critical point, $|\Psi|$ keeps a very small value.
After $\hat{t}$, it increases very rapidly and then it starts to oscillate.
As we show later, this is a typical behavior of $|\Psi|$.
We determined the time $\hat{t}$ as $|\Psi(\hat{t})|=2|\Psi|_{\rm critical point}$
as in the previous paper \cite{hol}.
In Fig.~\ref{OPPNAS}, we also indicate $\hat{t}$ and the time at which the measurement
of the correlation length was done in Ref. \cite{Braun}. 
The KZM and KZ scaling should be applied for the region near $\hat{t}$, whereas
the measurement was performed in the oscillating regime.
We think that this is the source of the above mentioned discrepancy.
That is, there exists at least two distinct temporal regimes in which 
the system in quench exhibits different dynamical behaviors.
In fact, the calculation of $|\Psi|$ in Fig.~\ref{OPPNAS} shows that there exists 
another important
time, which we call $t_{\rm eq}$, for the evolution of the BEC under a slow quench.
At time $t_{\rm eq}$, the BEC starts to oscillate, and it is expected that
a certain important coarsening process takes place after $t_{\rm eq}$. 
{\em We expect that the KZ scaling is only applicable for the regime from 
$\hat{t}$ to $t_{\rm eq}$}. 

In the subsequent sections, we shall reveal the dynamical behavior of lattice boson
systems in quantum quench and verify the above expectation.

\section{Evolution of BEC order parameters, vortex density and correlation
length for KZM protocol}{\label{KZMP}}

\begin{figure}[h]
\centering
\begin{center}
\includegraphics[width=6.2cm]{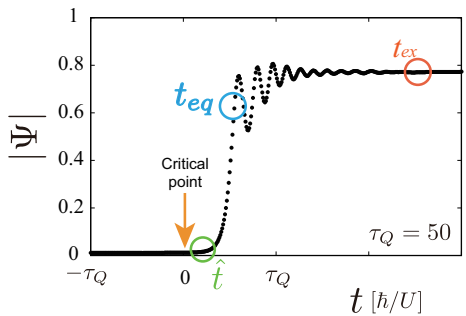}
\end{center}
\caption{Temporal evolution of the amplitude of the order parameter
$|\Psi|$ for the KZ protocol.
Critical point, $\hat{t}$ and $\teq$ are indicated.
In this protocol with $\tQ=50$, $\hat{t}\approx 14$ and $\teq \approx 27$.
}
\label{OPKZ}
\end{figure}
\begin{figure}[h]
\centering
\begin{center}
\includegraphics[width=6.2cm]{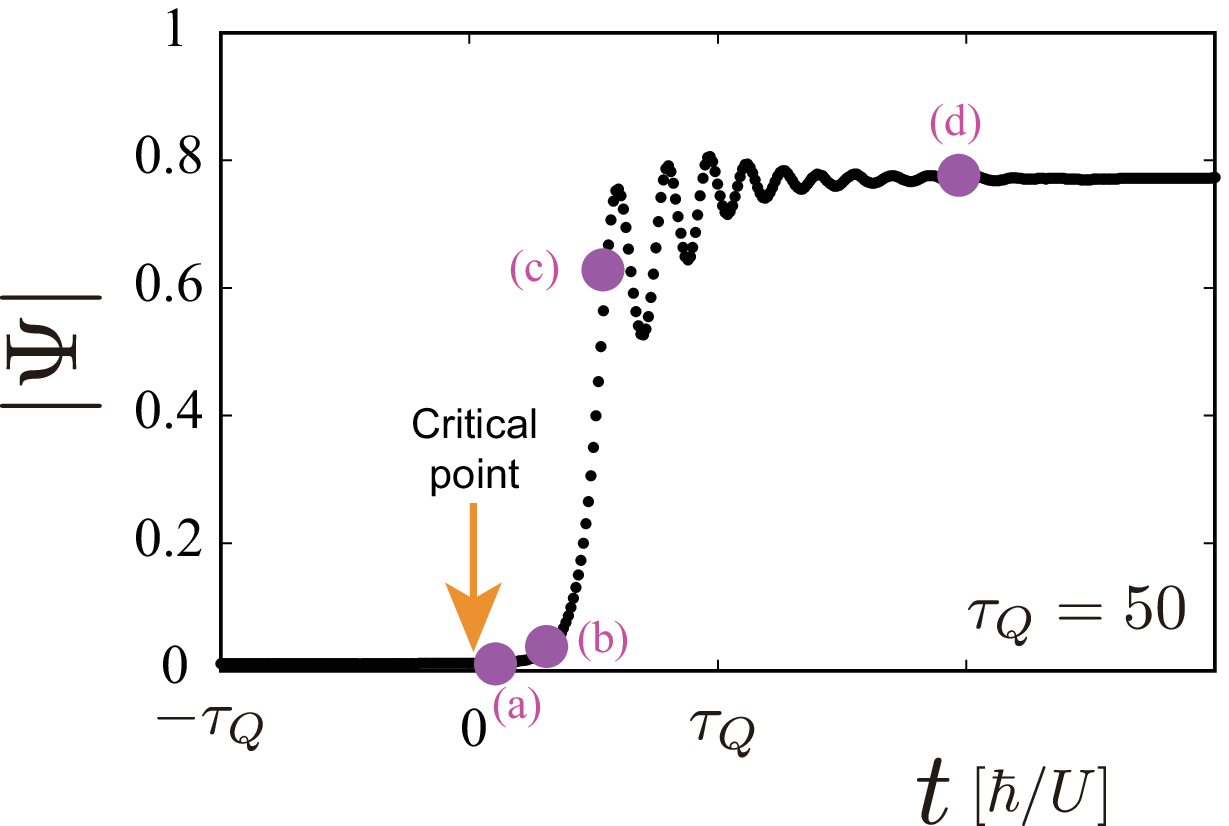}
\includegraphics[width=7.5cm]{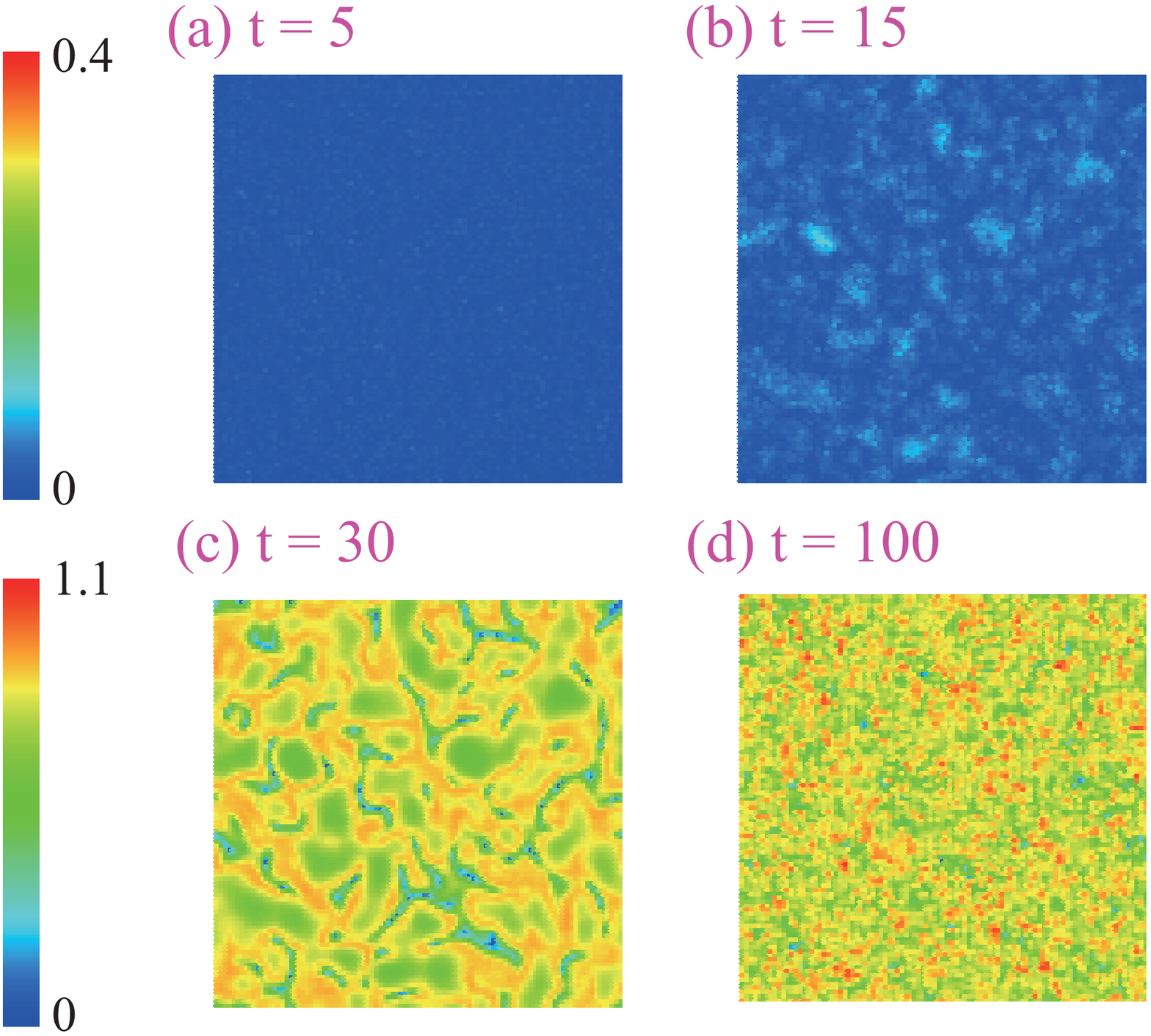}
\end{center}
\caption{Snapshots of $|\Psi_i|$ for various times (a)-(d) indicated in the upper
panel. 
SF local amplitude is getting larger.
Domain structure at $t=30$ tends to disappear at $t=100$.
Unit of time is $\hbar/U$. 
}
\label{SFdensity}
\end{figure}
\begin{figure}[h]
\centering
\begin{center}
\includegraphics[width=7.5cm]{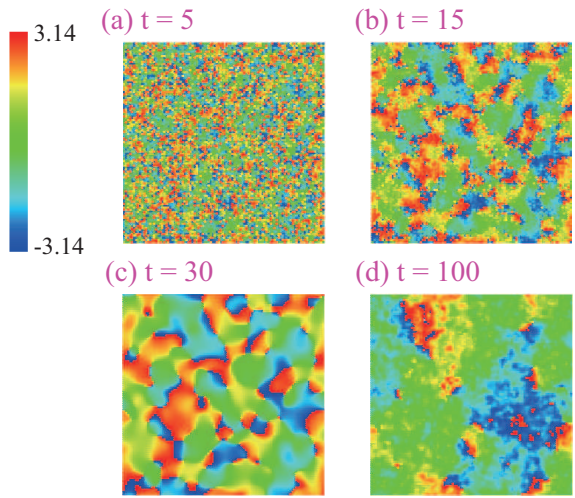}
\end{center}
\caption{Snapshots of phase of $\Psi_i$ for various times corresponding to 
Fig.~\ref{SFdensity}.
Size of domains is getting larger, but the domain structure remains even 
at the end of the oscillating behavior of the amplitude of $|\Psi|$ [$t=100$].
Unit of time is $\hbar/U$. 
}
\label{SFphase}
\end{figure}
\begin{figure}[h]
\centering
\begin{center}
\includegraphics[width=7.5cm]{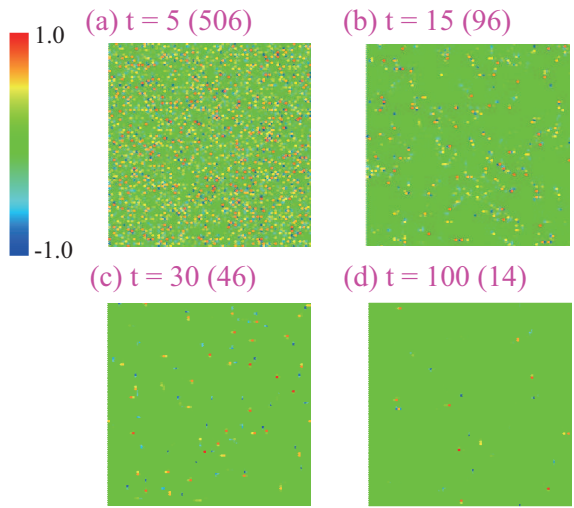}
\end{center}
\caption{Snapshots of vortex $\Omega_i$ for various times corresponding to 
Fig.~\ref{SFdensity}.
The numbers in the parentheses are the number of (vortex) + (anti-vortex),
i.e., $N_{\rm v}$.
Unit of time is $\hbar/U$. 
}
\label{Vortex}
\end{figure}

In the previous section, the possible source of the discrepancy between
the experimental results in Ref.~\cite{Braun} and the KZ scaling prediction
was pointed out.
In this section, we numerically study the dynamics of the Bose-Hubbard model
in the quench of the KZ protocol and see if the KZ scaling holds.
We put the initial time $t_{\rm i}=-\tQ$, and then from Eq.(\ref{protocol}), 
$J(t_{\rm i})=0$, i.e., the system is in the deep Mott insulator.
On the other hand, the quench terminates at $t_{\rm f}=\tQ$,
i.e., $J(t_{\rm f})=2J_{\rm c}$.
The system size is $100\times 100$ for most of our numerical studies,
whereas we have verified system-size dependence when we judged its necessity.

The calculated BEC amplitude $|\Psi|$ is shown in Fig.~\ref{OPKZ}, 
which exhibits a similar behavior to that in Fig.~\ref{OPPNAS}.
In Fig.~\ref{SFdensity}, we show typical snapshots of the local amplitude $|\Psi_i|$.
Just after passing cross the critical point, $|\Psi|$ is very small, whereas
from $t=\hat{t}$ to $t=t_{\rm eq}$, $|\Psi|$ develops very rapidly.
We notice that $t_{\rm eq}/\hat{t} \approx 2$, and therefore this period
is rather short.
However, $|\Psi(t_{\rm eq})|/|\Psi(\hat{t})|\approx 30$.
It is obvious that the clear domain structure forms at $t=30$
and the sizes of domains are rather large.
On the other hand after the oscillating behavior, a homogeneous state of
$|\Psi_i|$ forms at $t=100$. 

\begin{figure}[t]
\centering
\begin{center}
\includegraphics[width=6cm]{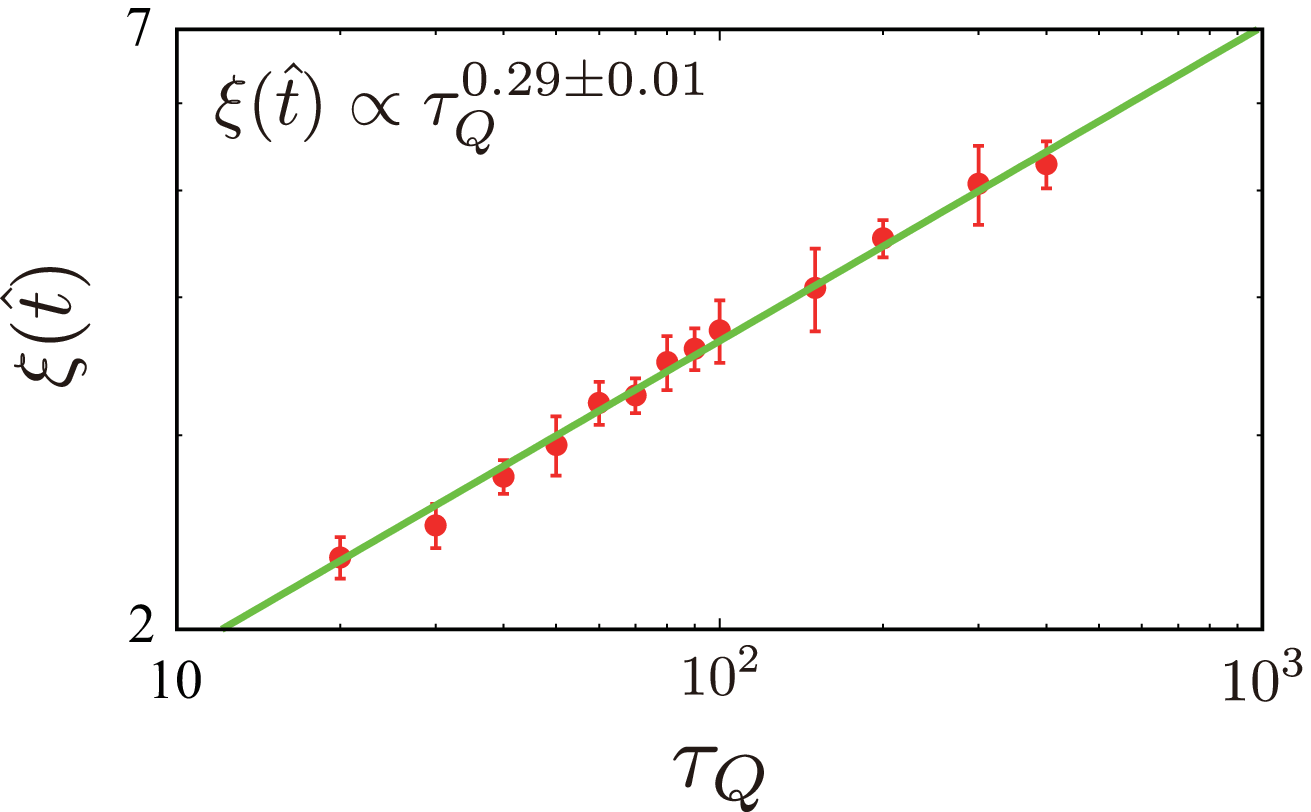}
\includegraphics[width=6cm]{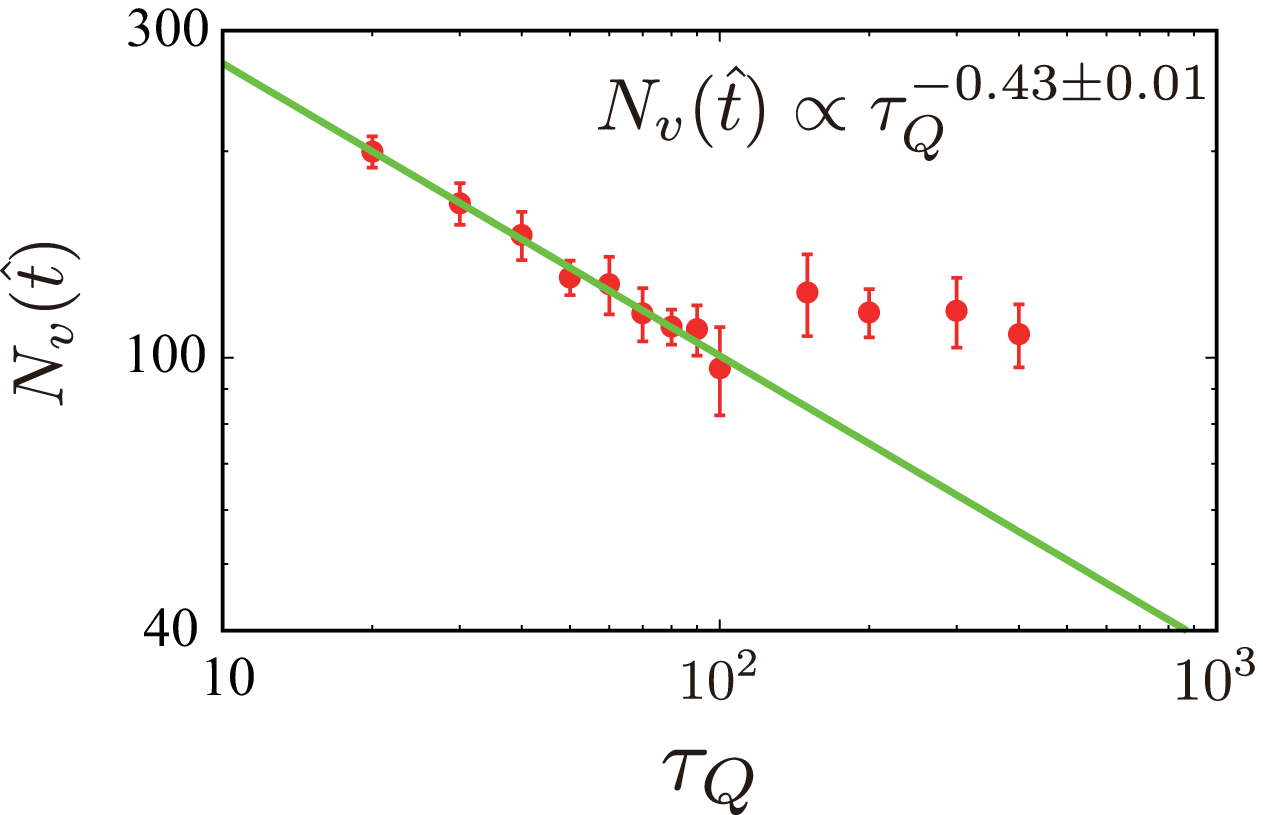}
\end{center}
\caption{Exponents of the correlation length $b$ and vortex density $d$
at $t=\hat{t}$.
Correlation length $\xi(\hat{t})$ exhibits a clear scaling law for $\tQ=20\sim 400$,
whereas vortex density $N_{\rm v}(\hat{t})$ shows anomalous behavior 
for $\tQ>100$.
Source of this behavior is discussed in the text.}
\label{exp1}
\end{figure}
\begin{figure}[h]
\centering
\begin{center}
\includegraphics[width=6cm]{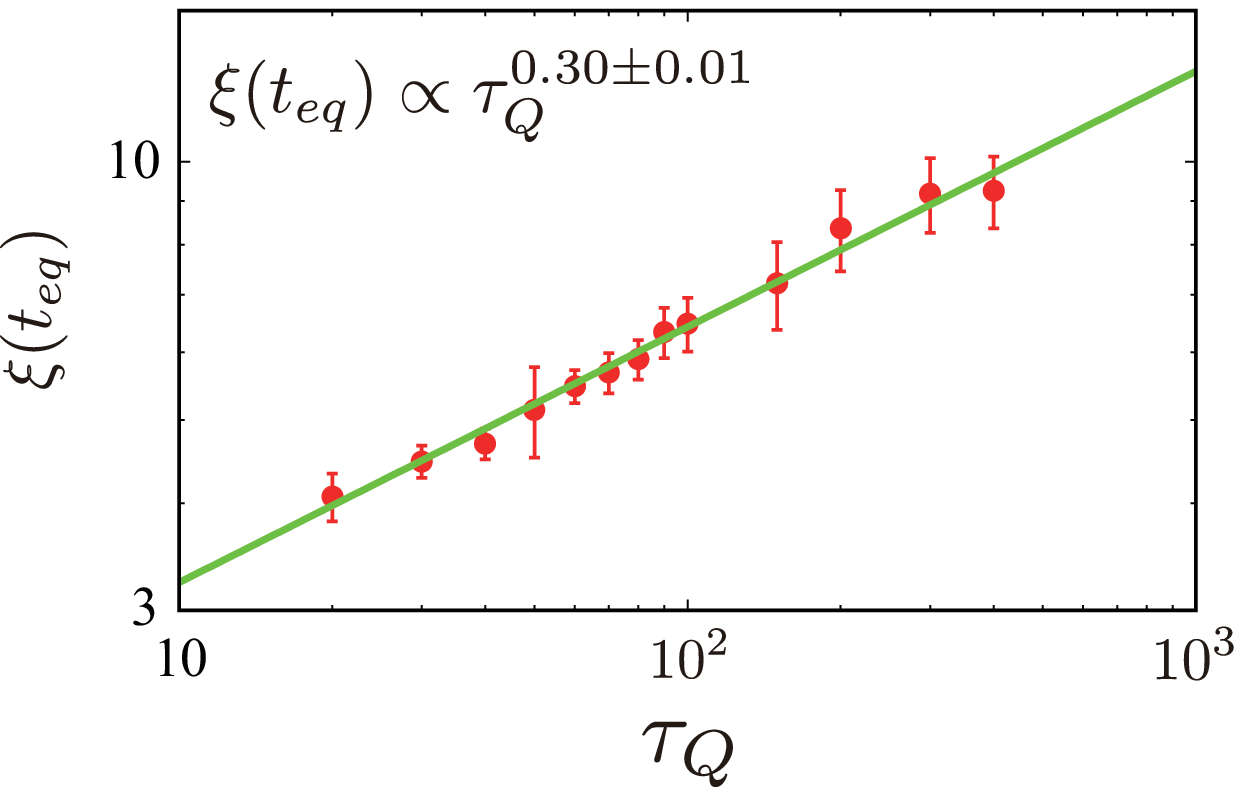}
\includegraphics[width=6cm]{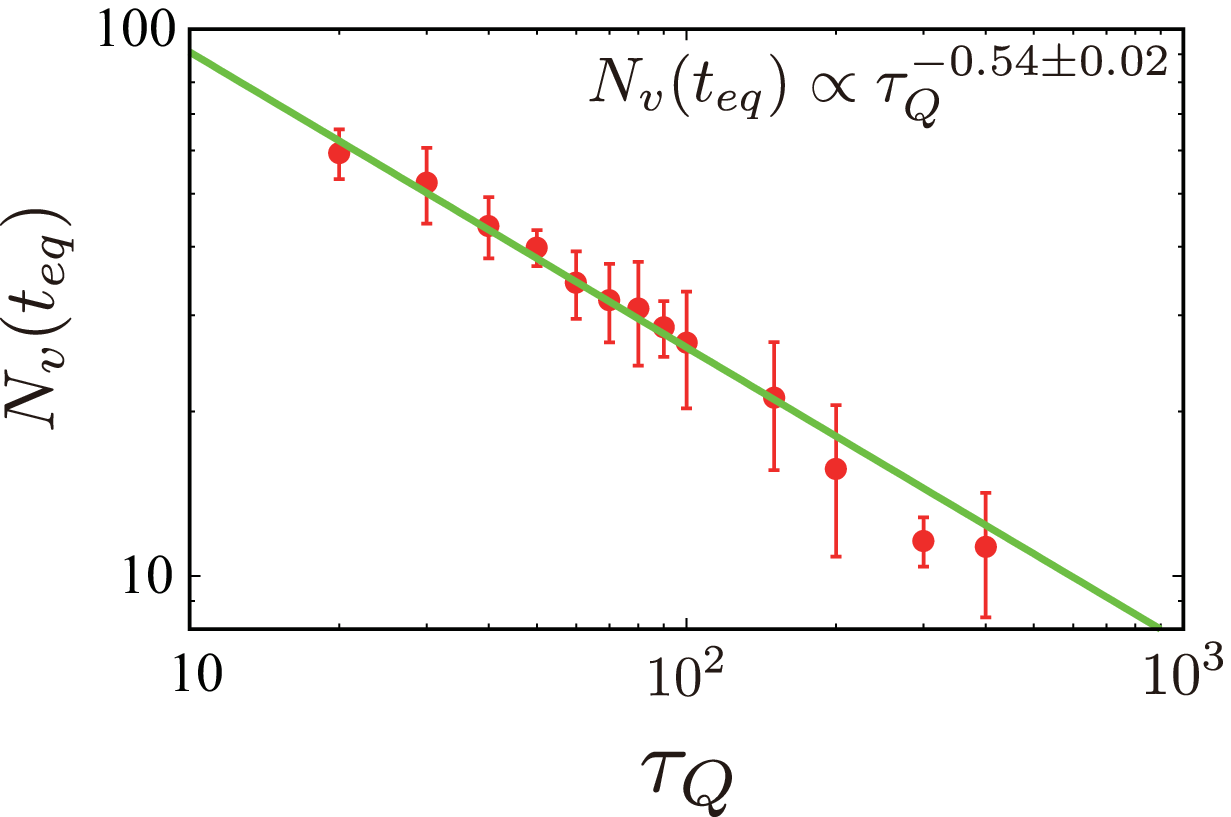}
\end{center}
\caption{Exponents of the correlation length $b$ and vortex density $d$
at $t=\teq$.
Both correlation length $\xi(\teq)$ and vortex density $N_{\rm v}(\teq)$
exhibit a clear scaling law for $\tQ=20\sim 400$.
For $\tQ\leq 100$, the exponents $b$ and $d$ have close values with those at $t=\hat{t}$.}
\label{exp2}
\end{figure}
\begin{figure}[h]
\centering
\begin{center}
\includegraphics[width=6cm]{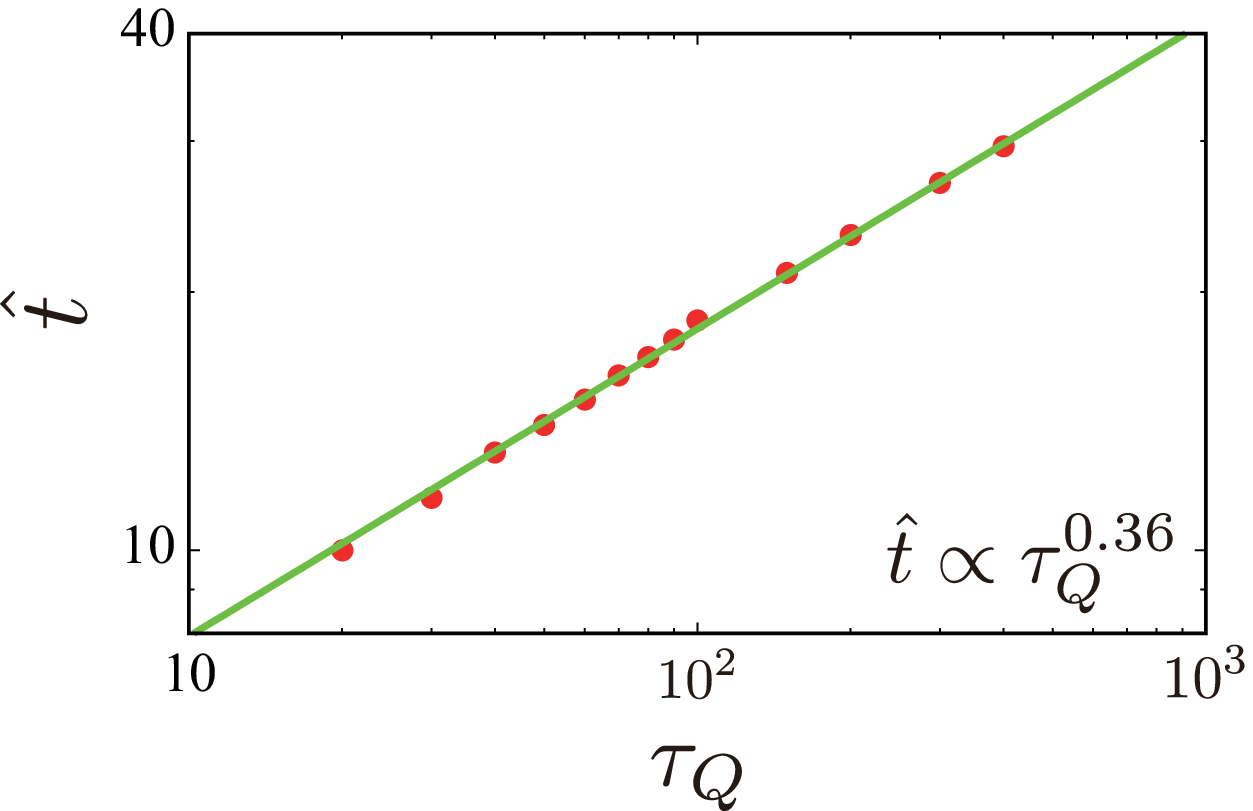}
\includegraphics[width=6cm]{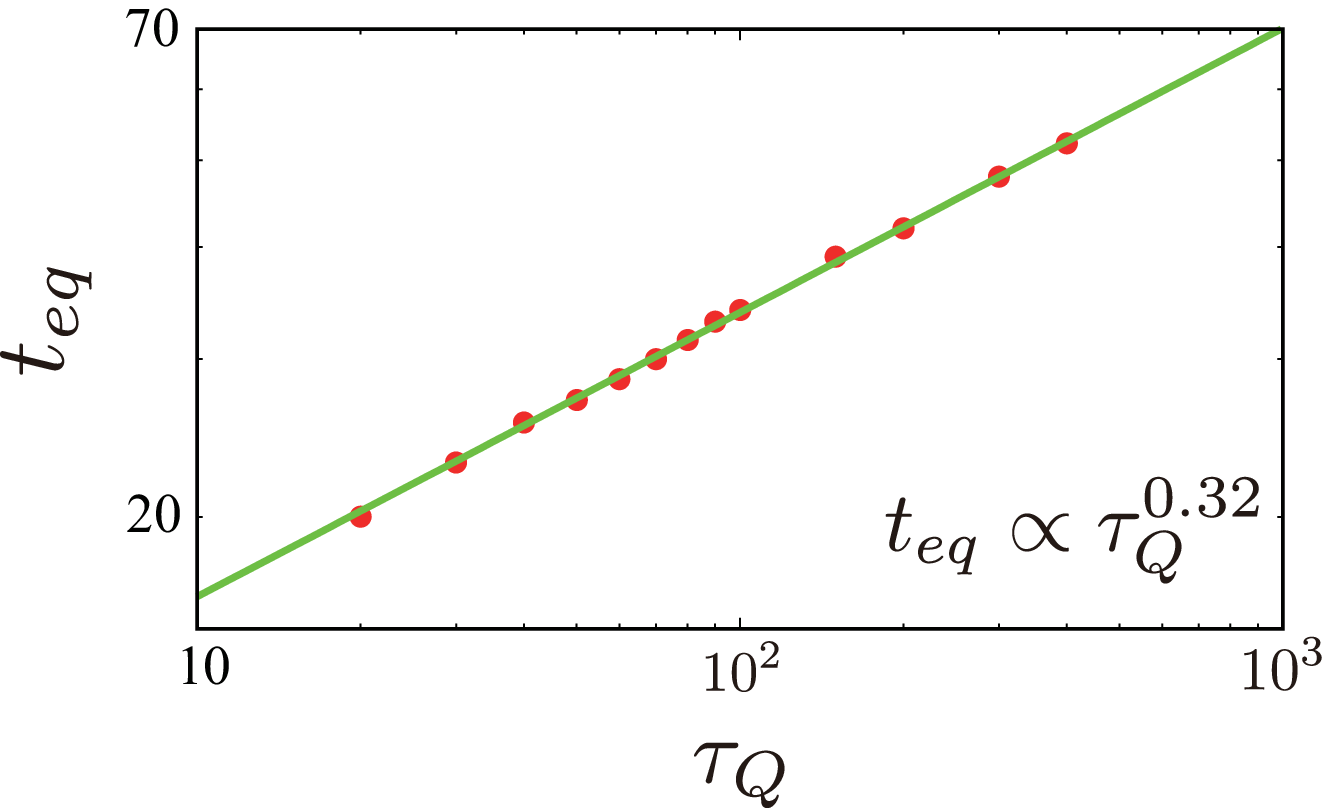}
\end{center}
\caption{Scaling law of $\hat{t}$ and $\teq$ with respect to $\tQ$.
Both $\hat{t}$ and $\teq$ satisfy almost the same scaling.
Therefore, $\hat{t}(\tQ)\propto \teq(\tQ)$.}
\label{exp3}
\end{figure}

It is also interesting to see phases of $\Psi_i$.
See Fig.~\ref{SFphase}.
Just after the critical point, small domains form, and for $t=15-100$, 
sizes of domains are getting larger and the local amplitude $|\Psi_i|$ 
and phase of the BEC have apparent correlations for $t=30$.
We have verified that similar correlations between the local density and phase
of the SF order parameter exist for other $t(<100)$.
After the amplitude $|\Psi|$ terminates the oscillation at $t=100$, 
there still exists a domain structure in the phase of $\Psi_i$.
This makes the correlation length finite although the order parameter has 
a large amplitude.

We also show the vortex configurations and the number of vortex in Fig.~\ref{Vortex}.
At $t=15$, a substantial decrease of vortices has already taken place from 
the moment passing across the critical point.

From the above calculations, we expect that somewhat smooth evolution 
of the phase of the BEC takes place between $\hat{t}$ and $\teq$ although 
its amplitude $|\Psi|$ increases rapidly.
In fact,
this expectation is supported by the calculations of the correlation length and the
vortex density at $\hat{t}$ and $\teq$, i.e., we have $\xi(\teq)/\xi(\hat{t})\approx 2$ 
and $N_{\rm v}(\teq)/N_{\rm v}(\hat{t})\approx 0.3$.
This observation is one of the key points for understanding dynamics and coarsening 
process for the SF formation in the Bose-Hubbard model as we shall show.
It should be remarked here that the evolution of the system is a 
non-adiabatic and off-equilibrium phenomenon from $\hat{t}$ to $\teq$, 
and also in the oscillation period.

In order to see if the KZ scaling hypothesis holds,
we calculated the exponents $b$ for the correlation length $\xi\propto \tQ^b$
and $d$ for $N_{\rm v}\propto \tQ^{-d}$ both at $t=\hat{t}$ and $t=\teq$
for various $\tQ$ [$20\leq \tQ\leq 400$].
If $b$ and $d$ have close values at $t=\hat{t}$ and $t=\teq$, the smooth
evolution picture of the BEC between $\hat{t}\sim\teq$ has further
supports.
We show the results in Figs.~\ref{exp1} and \ref{exp2}.
From $\tQ=20$ to $100$, the correlation length and vortex density both exhibit
scaling law and the exponents for the data at $t=\hat{t}$ and $t=\teq$ are close
with each other.
For $\tQ>100$, the vortex density $N_{\rm v}(\hat{t})$ shows anomalous behavior
although the other quantities satisfy the scaling law up to 400.
We have verified that this anomalous behavior of  $N_{\rm v}(\hat{t})$ is {\em not}
a finite-size effect by calculating $N_{\rm v}(\hat{t})$ for the system size
$(64\times 64)$.
To understand this result, we show the values of $J(\hat{t})$ for various $\tQ$.
See Table~\ref{valueJ}. 
As $\tQ$ is getting larger, $J(\hat{t})$ is getting closer to the critical point.
Just after passing through the critical point, a large number of vortices exist
and as a result, a proper scaling law does not hold.

\begin{table}[h]
\caption{ The values of $J(\hat{t})$ for various $\tau_Q$.
The critical value of the Mott-SF phase transition is $J_{\rm c}=0.043$.}
  \begin{tabular}{|c||c|c|c|c|c|c|c|} \hline
    $\tau_Q$     & 20    & 40    & 60  & 80  & 100   &200 & 400   \\ \hline
    $J(\hat{t})$ & 0.065 & 0.057 &0.054&0.052& 0.051 &0.048&  0.046\\ \hline
  \end{tabular}
\label{valueJ}
\end{table}

The exponents $b$ and $d$ are estimated as $b\approx 0.3$ and
$d\approx  (0.45-0.55)$.
It should be remarked that these values are not in agreement with those obtained
by the exponents of the 3D XY model ($b=0.402, \ d=0.804$), nor
those by the mean field theory ($b=0.25, \ d=0.5$ from $\nu=1/2$ and $z=2$), 
but they are rather in-between. 
Critical exponents of the Bose-Hubbard model {\em obtained by the 
Gutzwiller methods}
might be different from those of the simple mean-field-theory. 
This problem is under study and we hope that the results will be reported
in the near future.

\begin{figure}[h]
\centering
\begin{center}
\includegraphics[width=4cm]{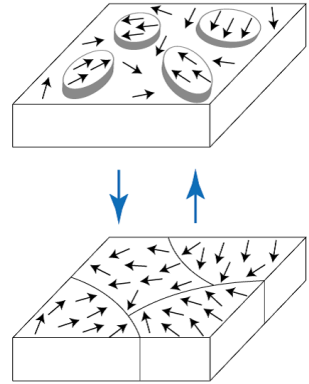}
\end{center}
\caption{Physical picture of the coarsening process from $\teq$ to $\tex$.
Dense regions in an inhomogeneous state have robust BEC, whereas
a homogeneous state has large SF flows by phase-ordering process.}
\label{coars}
\end{figure}

In order to certify the similarity of the states at $t=\hat{t}$ and $t=\teq$
for $\tQ\leq 100$, we furthermore investigated the scaling law of  $t=\hat{t}$ 
and $t=\teq$ with respect to $\tQ$.
The results are shown in Fig.~\ref{exp3}.
It is obvious that $t=\hat{t}$ and $t=\teq$ have very close exponents, i.e.,
$\hat{t}(\tQ)\propto \teq(\tQ)$.
It is interesting that the scaling law holds up to $\tQ=400$, the maximum value
of the present numerical study.

From the above consideration, we conclude that from $t=\hat{t}$ to $t=\teq$, 
smooth evolution of the phase degrees of freedom of the BEC and
the topological defects, i.e., vortices, takes place although the rapid
increase of the amplitude of the order parameter occurs there.
That is, from $t=\hat{t}$ to $t=\teq$, the system in the quench experiences 
a smooth coarsening (phase-ordering) process.
In Ref.~\cite{sondhi}, the scaling hypothesis, which states that all observables depend
on a time $t$ through $t/\hat{t}$ and on a distance $x$ through $x/\xi(\hat{t})$,
was proposed.
It is naturally expect that this hypothesis holds in the time region from $\hat{t}$
to $\teq$.

On the other hand as Fig.~\ref{SFphase} shows, a strong coarsening process 
occurs in the amplitude oscillating regime after $t>\teq$.
We measured the time evolution of the average kinetic and on-site interaction 
energies separately, and found that they synchronize with the behavior of the BEC.
When the amplitude of the BEC is large, the kinetic (interaction) energy is small (large),
whereas when the amplitude of the BEC is small, the kinetic (interaction) energy is 
large (small).
From these observation, we have got a physical picture of the coarsening process
such that the large amplitude of the BEC comes from an inhomogeneous particle
distribution and the small amplitude state accompanies a SF flow to make 
the system homogeneous.
As a result, domains are getting larger in the oscillating regime.
See Fig.~\ref{coars}.

\section{Behavior after quench}{\label{after}}

In the previous section, we observed the dynamics of the BEC after 
passing through the critical point, and have obtained the physical picture
of the evolution.
The oscillating behavior of the BEC is getting weak and finally terminates 
at certain time, which we call $t_{\rm ex}$.
However, the phase of the order parameter of the BEC is still not uniform
and topological defects exist at $t=t_{\rm ex}$.
Therefore one may wonder to which state the BEC approaches after
the quench.
This is an important problem from the view point of the statistical mechanics,
i.e., whether an equilibrium state forms or inhomogeneous amorphous like 
state persists \cite{polkovnikov}.
As the system after the quench is an isolated system and therefore
the total energy is conserved.
Then, both of the above two scenarios are possible.

\begin{figure}[h]
\centering
\begin{center}
\includegraphics[width=6cm]{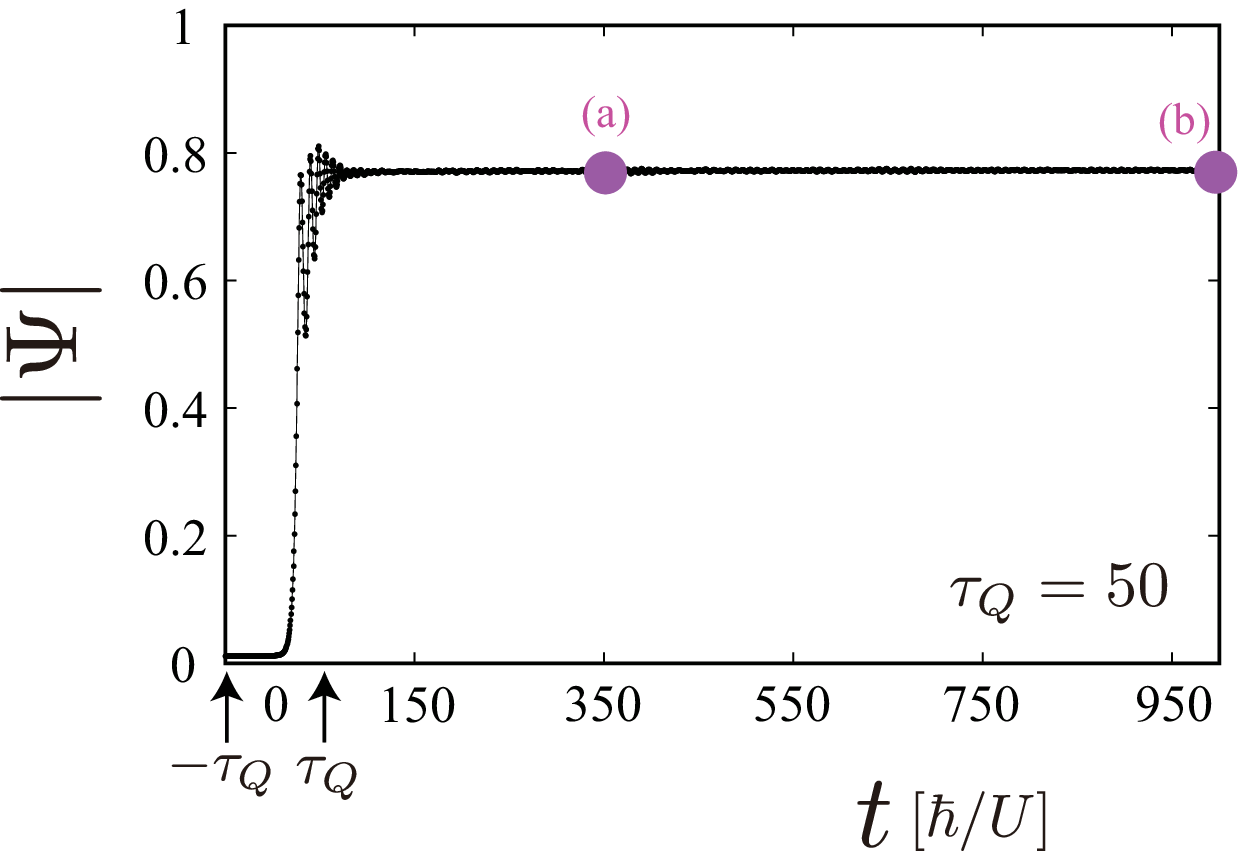}
\includegraphics[width=6.5cm]{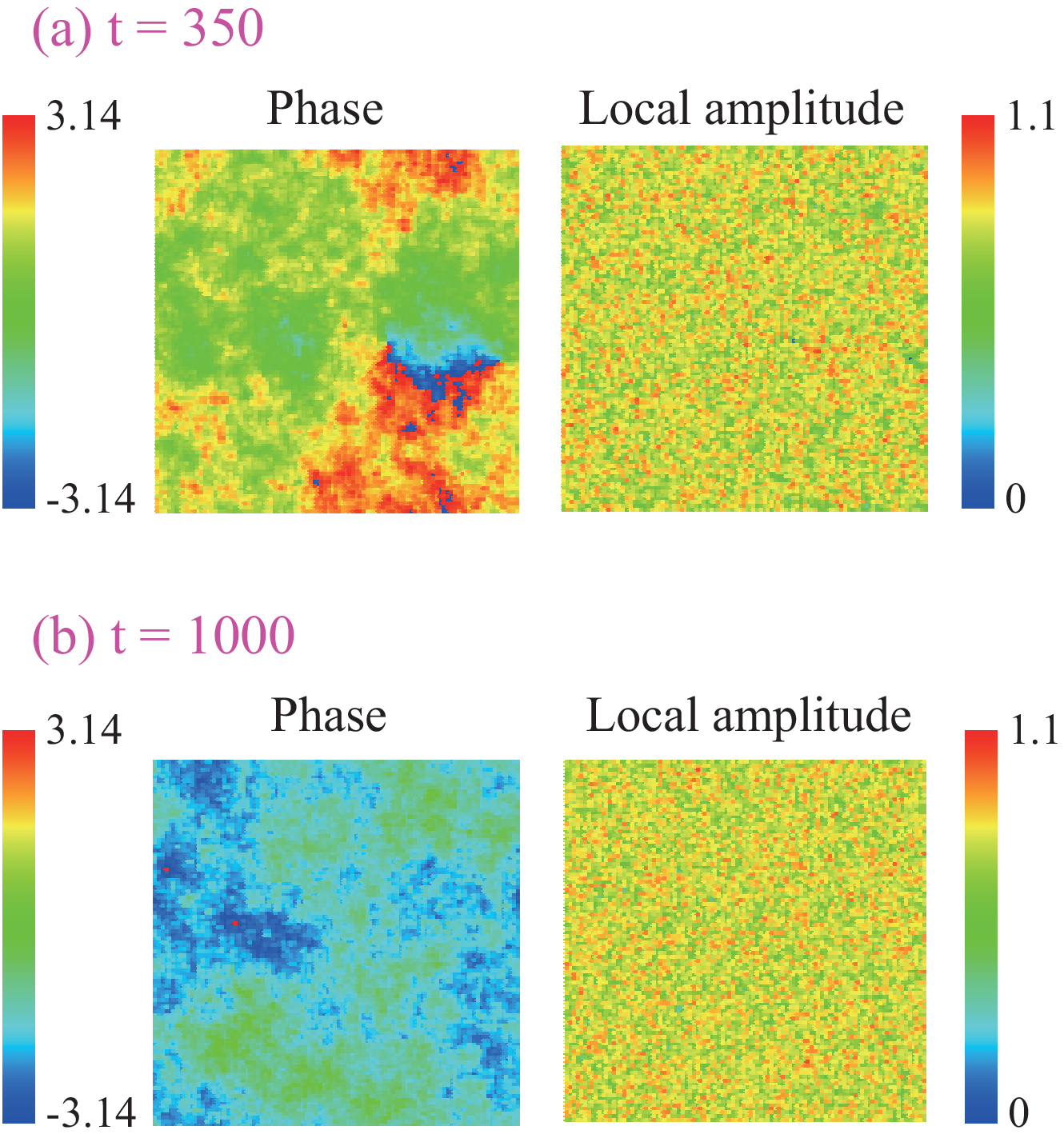}
\end{center}
\caption{Long-time behaviors of the order parameter of BEC.
$\tQ=50$.
}
\label{LT}
\end{figure}

In order to study the above problem, we observed the long-time behavior of the system
after the quench.
Local amplitude $|\Psi_i|$ and phase degrees of freedom were calculated
for a long period and the results are shown in Fig.~\ref{LT}.
Local amplitude $|\Psi_i|$ is rather homogeneous already in the intermediate regime,
as we observed in the previous section.
On the other hand, the phase of BEC has a domain structure for a rather long
period, whereas the system finally results in a fairly uniform state.
In fact, the state at $t=1000$ in Fig.~\ref{LT} does not have topological vortices.
(Small but finite fluctuations in the phase of the BEC are inevitable even in the SF.)
The coarsening process to this state might be understood by the phase-ordering 
study done in 1990's \cite{PO}.
As the uniform BEC is expected to have a lower energy that the amorphous like state,
some amount of energy might have transfered to the particles that are not participating
the BEC.
These problems are under study and we hope that results will be published
in the near future.


\section{Conclusion and discussion}{\label{conclusion}}

In this paper, we have obtained the global picture of the dynamics of the quantum
phase transition from the Mott insulator to SF by using the tGW methods.
We focused on dynamical quantum phase transition in the 2D system, for which
it is difficult to apply state-of-the art numerical simulations such as the
time-dependent-density-matrix-renormalization group and
time-evolving-block decimation.
We first examined the recent experimental measurements of the exponent 
of correlation length 
in Ref.~\cite{Braun}, and showed that our numerical study gives almost the same
values of the exponent.
We also pointed out the possible source of the discrepancy between the KZ scaling
hypothesis and the above experimental and numerical results.

Then, we studied dynamical behavior of the Bose-Hubbard model in the quench
by solving the time-dependent Schr\"{o}dinger equation by the tGW methods.
We found that the amplitude of the BEC exhibits the interesting behavior
and there exist other important time scales $\teq$ and $\tex$, besides $\hat{t}$. 
The existence of the time scale $\teq$ was discussed in Ref.~\cite{hol},
and the amplitude of the BEC develops very rapidly from $\hat{t}$ to $\teq$.
On the other hand from $\teq$ to $\tex$, the BEC amplitude oscillates 
rather strongly.
In order to understand these behaviors, we calculated the local amplitude,
local phase of the BEC order parameter, and vortex density.
From $\hat{t}$ to $\teq$, rather smooth coarsening process takes place
in spite of the large increase in the BEC amplitude.
In contrast to this period, from $\teq$ to $\tex$, hard coarsening process
occurs.
The phase-ordering process accompanies local fluctuations of the particle
density as well as the BEC order parameter. 

In order to study whether the KZM-like scaling holds in the quench dynamics, 
we calculated the correlation length and vortex density as a function of 
the quench time $\tQ$.
We obtained the exponents $b$ and $d$ and found that these have very close
values at $t=\hat{t}$ and $\teq$.
This result supports the physical picture such that the evolution of the BEC
from $\hat{t}$ to $\teq$ is rather smooth from the view point of the 
phase-ordering coarsening process.
Another important observation that we have is that for very slow quench
$\tQ\to \infty$, the vortex density at $t=\hat{t}$ does not satisfy a simple
scaling law.
This comes from the fact that $\hat{t}$ is getting large but simultaneously
$J(\hat{t})\to J_{\rm c}$ in this case.
For experimental setups, this observation might be useful.

Finally, we observed how the system evolves after $\tex$ as the state after the 
quench has an amorphous-like domain structure.
We found that this domain structure remains rather long time but at the end 
the system settled in a uniform state and we think that this state is an equilibrium 
state.

It is useful to comment on the experimental setup to observe the new findings
in the present work. 
In particular for experiments on ultra-cold atomic gases, 
it is useful to show tuning conditions of the on-site interaction energy $U$.
In the experimental setups,
a typical upper limit of the holding time in practical experiments is $100$[ms] \cite{Will}. 
$U$ is changed by Feshbach resonance technique. 
As an example,
Table.\ref{time-scale_atom} shows the tuning condition of $U$ 
for recent experimental systems with ${}^{39}$K and ${}^{87}$Rb cold atoms.

\begin{table}[t]
\caption{ Tuning of on-site interaction in the case $t=1\to 0.1$[ms]  in recent experimental systems.
$\lambda$ is wavelength of an optical lattice. $E_{R}$ is recoil energy, $E_{R}=h^{2}/(2m\lambda^{2})$ ($m$ is atom mass). }
  \begin{tabular}{|c|c|c|c|} \hline
                                         & $\lambda$ [nm] & $E_{R}/h$ [kHz] & $U$ [$E_{R}$]  \\ \hline
    Ref.\cite{Braun} ${}^{39}$K  &    736       &   9.4     &    0.17           \\ \hline
    Ref.\cite{Will}  ${}^{87}$Rb  &   738      &    4.2    &     0.38          \\ \hline
    Ref.\cite{Chen}    ${}^{87}$Rb  &    812      &    3.5    &     0.45          \\ \hline

 \end{tabular}
\label{time-scale_atom}
\end{table}

\section*{Acknowledgments}
We thank Y. Takahashi and Y. Takasu 
for helpful discussions from the experimental point of view.
Y. K. acknowledges the support of a Grant-in-Aid for JSPS
Fellows (No.17J00486).



\begin{thebibliography}{99}
\bibitem{initial}
For the Mott initial states, we set relative phases of $\{{f^{i}_{n}(t_{\rm i})}\}$ 
fully-random and also applied small local fluctuations in $\{\langle n_{i}(t_{\rm i})\rangle\}$. 
This is the origin of the quantum-statistical fluctuations in the tGW calculation.
\end{thebibliography}
\end{document}